\documentclass[conference]{IEEEtran}
\IEEEoverridecommandlockouts
\usepackage{cite}
\usepackage{amsmath,amssymb,amsfonts}
\usepackage{algorithmic}
\usepackage{graphicx}
\usepackage{textcomp}
\usepackage{xcolor}
\usepackage{listings}
\usepackage{hyperref}
\usepackage{xcolor} 
\hypersetup{
    colorlinks = true,    
    linkcolor  = blue,    
    citecolor  = blue,    
    urlcolor   = blue     
}

\begin{document}

\def\BibTeX{{\rm B\kern-.05em{\sc i\kern-.025em b}\kern-.08em
    T\kern-.1667em\lower.7ex\hbox{E}\kern-.125emX}}

\title{Automatic Qiskit Code Refactoring Using Large Language Models}

\author{\IEEEauthorblockN{1\textsuperscript{st} José Manuel Suárez}
\IEEEauthorblockA{\textit{LIFIA} \\
\textit{UNLP}\\
La Plata, Buenos Aires, Argentina \\
jsuarez@lifia.info.unlp.edu.ar}
\and
\IEEEauthorblockN{2\textsuperscript{rd} Luis Mariano Bibbó} \IEEEauthorblockA{\textit{LIFIA} \\
\textit{UNLP}\\
La Plata, Buenos Aires, Argentina \\
lmbibbo@lifia.info.unlp.edu.ar}
\and
\IEEEauthorblockN{3\textsuperscript{nd} Joaquin Bogado}
\IEEEauthorblockA{\textit{LIFIA} \\
\textit{UNLP}\\
La Plata, Buenos Aires, Argentina \\
jbogado@lifia.info.unlp.edu.ar}
\and
\IEEEauthorblockN{4\textsuperscript{th} Alejandro Fernandez}
\IEEEauthorblockA{\textit{LIFIA} \\
\textit{UNLP}\\
La Plata, Buenos Aires, Argentina \\
alejandro.fernandez@lifia.info.unlp.edu.ar}
}

\maketitle

\begin{abstract}
As quantum software frameworks evolve, developers face increasing challenges in maintaining compatibility with rapidly changing APIs. In this work, we present a novel methodology for refactoring Qiskit code using large language models (LLMs). We begin by extracting a taxonomy of migration scenarios from the different sources of official Qiskit documentation (such as release notes), capturing common patterns such as migration of functionality to different modules and deprecated usage. This taxonomy, along with the original Python source code, is provided as input to an LLM, which is then tasked with identifying instances of migration scenarios in the code and suggesting appropriate refactoring solutions. Our approach is designed to address the context length limitations of current LLMs by structuring the input and reasoning process in a targeted, efficient manner. The results demonstrate that LLMs, when guided by domain-specific migration knowledge, can effectively assist in automating Qiskit code migration. This work contributes both a set of proven prompts and taxonomy for Qiskit code migration from earlier versions to version 0.46 and a methodology to asses the capabilities of LLMs to assist in the migration of quantum code.
\end{abstract}
\date{May 2025}
\maketitle

\section{Introduction}
The rapid evolution of quantum software frameworks \cite{preskill_quantum_2018} presents a unique challenge for developers maintaining code across versions \cite{jimenez-navajas_code_2025}. This issue is particularly evident in Qiskit, one of the most widely adopted platforms for quantum programming. The recent release of version 2.0 introduced substantial changes that affect backward compatibility. Major releases, typically published on an annual basis, often include significant modifications ranging from API deprecations to more fundamental architectural updates. As a result, developers must invest considerable effort in understanding and adapting their existing code bases to align with the latest version.

Migrating code between Qiskit versions is not only time-consuming but also error-prone \cite{asif_pennylang_2025, javadi-abhari_quantum_2024, dupuis_qiskit_2024, nation_benchmarking_2025}, especially for developers who are not deeply familiar with the internal evolution of the framework. Developers often find their programs broken after library updates \cite{li_instructcoder_2024, huynh_survey_2025, khojah_impact_2024}, leaving them reading documentation and release notes to understand a new set of problems in once functional code \cite{tsantalis_refactoringminer_2022, quetschlich_mqt_2023}. This raises a key question: \textit{'Can large language models (LLMs) assist in the process of Qiskit code migration by leveraging structured knowledge about breaking changes?'}. On the other hand, as the Qiskit ecosystem evolves, new versions often introduce not only breaking changes but also enhancements aimed at improving performance, usability, and modularity of quantum algorithms. These changes, while beneficial, may not be immediately adopted by practitioners maintaining legacy code or developing new applications based on outdated paradigms. Consequently, beyond simply ensuring compatibility, there is a growing need to leverage these improvements to optimize existing quantum software. This raises a compelling question in the context of Quantum Software Engineering (QSE): \textit{can large language models (LLMs), when equipped with domain-specific migration knowledge, go beyond syntactic adaptation and suggest substantive improvements that reflect best practices and leverage newly introduced functionalities?}

In this work, we explore a novel methodology for refactoring Qiskit code using LLMs guided by a domain-specific taxonomy of migration scenarios. The taxonomy, developed in Suárez et. al~\cite{previous_paper}, was created from Qiskit’s official documentation through both manual and LLM-assisted processes. For the version in our experiment, it consists of approximately 43 representative cases, covering categories such as deprecations, new features, module restructuring, renamed classes, parameter changes, and architectural shifts. Compared to raw documentation, the taxonomy offers a condensed and structured representation of migration knowledge that is more readily digestible by both humans and LLM models.

Our method involves providing an LLM with the migration taxonomy, a synthetic Python source file in a known version of Qiskit library and has a known set of migration scenarios present, and a prompt asking the model to identify instances of the migration scenarios within the code. This step allow us to evaluate the performance of the model to identify the migration scenarios present in the taxonomy. In a later stage, the LLM is also asked to propose migrated versions of the identified segments. To assess the feasibility and accuracy of this approach, we conducted two sets of experiments: the first focused on identifying scenario instances in real-world and synthetic code examples; the second evaluated the quality of the migration suggestions on synthetic inputs. In both cases, the results were manually reviewed for correctness and relevance, and the model was required to explicitly reference the applicable migration scenario from the taxonomy.

While the use of LLMs for code transformation is becoming increasingly common \cite{dupuis_qiskit_2024, baumgartner_ai-driven_2024, yuan_transagent_2024}, our work distinguishes itself by anchoring the model's reasoning in a structured, domain-specific migration scenarios described in the taxonomy. This enables more targeted analysis than general-purpose prompting alone. Furthermore, thanks to recent advances in LLM context length, we are able to input both the taxonomy and source files in a single prompt, eliminating the need of advanced techniques, like RAG or elaborate chunking strategies.

The contributions of this work are twofold: (1) we present a novel methodology for combining domain-specific migration knowledge with LLM capabilities to refactor Qiskit code, and (2) we provide experimental evidence demonstrating that LLMs can effectively identify and resolve migration issues in a structured and interpretable way, when provided with structured domain specific knowledge.

The remainder of this paper is organized as follows: Section II reviews related work on software migration and LLM-based code transformation. Section III describes the Qiskit migration taxonomy and experimental setup. Section IV presents our methodology. Section V present the results and insights from our evaluation. Finally, Section VI concludes with a discussion of the results and directions for future work.

\section{Related work}
\label{sec:relatedwork}
The increasing complexity of quantum software, coupled with the rapid evolution of development frameworks like Qiskit, has exposed serious challenges in code maintenance, especially for tasks such as API migration and adaptation across versions. Large Language Models (LLMs) are emerging as versatile tools for assisting with these tasks, offering potential solutions in program explanation, generation, and transformation. While prior studies have explored individual capabilities of LLMs, few have directly examined their utility in coping with evolving quantum SDKs—a critical bottleneck in the scalability and longevity of quantum software. Here we discuss the contributions that are most pertinent to and aligned with the objectives of this study.

Frequent API changes in libraries like Qiskit introduce semantic drift, deprecated constructs, and updated usage patterns that break existing code. Yet, tool support for automatic migration is virtually nonexistent. Almeida et al. \cite{almeida_automatic_2024} investigated the use of GPT-4 for library migration tasks in Python, showing that carefully designed prompts (e.g., one-shot and chain-of-thought) significantly improve correctness during SQLAlchemy \footnote{https://www.sqlalchemy.org/} version upgrades. Although their domain was classical, the methodology is highly relevant to quantum SDK migration, where breaking changes often go undocumented or under-specified. In contrast, our work focuses specifically on quantum programs and exposes how these prompt strategies perform under domain-specific constraints, such as quantum software engineering semantics.

A few recent efforts have begun to explore how LLM strategies can support quantum code migration. Asif et al. \cite{asif_pennylang_2025}, for example, present PennyLang, which uses LLMs to translate quantum programs from Pennylane to Qiskit, focusing on inter-framework migration. Their approach combines prompt engineering with retrieval techniques to align equivalent quantum operations across SDKs. In contrast, our work addresses the challenges of intra-framework migration within Qiskit itself, where semantic changes and API evolution introduce subtler, version-specific obstacles that require deep contextual understanding.

One promising direction for improving quantum code migration is Retrieval-Augmented Generation (RAG), introduced by Lewis et al. \cite{lewis2021retrievalaugmentedgenerationknowledgeintensivenlp}. RAG augments language models with a retrieval component that dynamically fetches relevant information—such as documentation or code examples—from an external corpus during inference. This architecture allows the model to generate more accurate and up-to-date responses, especially in domains where knowledge evolves rapidly, without requiring model retraining. In contrast to approaches like RAG, our study operates in a static, zero-shot setting. We evaluate LLMs based solely on their pre-trained parametric memory, without any retrieval or fine-tuning to expose how well they handle quantum code migration tasks when relying solely on internalized knowledge. 

Refactoring-oriented studies, such as that by Cordeiro et al. \cite{cordeiro_empirical_2024}, show that LLMs tend to outperform human developers when dealing with systematic code smells, yet often fall short in more context-dependent scenarios. This limitation is particularly relevant to code migration, where understanding nuanced and version-specific changes is critical. Our work extends this perspective by focusing on the correctness and clarity of explanations in version-sensitive quantum code, highlighting how LLMs can misinterpret logic due to undocumented or implicit semantic shifts across different Qiskit versions.

Zhao’s catalog of quantum-specific refactorings \cite{zhao_refactoring_2023} emphasizes the need for automated transformation tools that respect quantum constraints like entanglement and measurement, which are often altered across Qiskit versions. Our study brings an orthogonal contribution by highlighting how LLMs struggle to explain these quantum-specific patterns when they appear in unfamiliar or legacy code structures, thereby underlining a prerequisite for safe refactoring.

Although migration requires transformation, it relies on a deep understanding of both the legacy and target code. D’Aloisio et al. \cite{daloisio_exploring_2024} explored LLMs’ ability to explain quantum algorithms in OpenQASM, demonstrating consistent explanation quality under constrained contexts. Suárez et al. \cite{previous_paper} extend their work by including newer LLMs (Qween, Deepseek, GPT-4), broader datasets, and a more comprehensive evaluation protocol that includes Qiskit-specific constructs and prompt variations across versions. We follow up this work using the generated taxonomies to enrich the LLM context with domain specific information to identify migration scenarios between Qiskit versions.

Dupuis et al. \cite{dupuis_qiskit_2024} fine-tuned language models on Qiskit code to optimize quantum-specific code generation, achieving state-of-the-art results on the Qiskit HumanEval benchmark. In contrast to their focus on model specialization for generative tasks, our work targets support for manual refactoring by comparing our model’s performance—using a migration case taxonomy—with that of general-purpose models without domain-specific fine-tuning, to propose useful transformations in response to Qiskit’s syntactic and semantic evolution.

At the intersection of quantum computing and migration, we also find the work by Zhao \cite{zhao_refactoring_2023}, which addresses the challenges of refactoring quantum programs written in Q\#. Although primarily centered on maintainability and efficiency rather than migration across versions, it does propose a useful catalog of refactorings based on algorithmic patterns — conceptually similar to our taxonomy. Furthermore, much like their proposed QSharp Refactoring Tool, our work moves toward building a hybrid tool for partially automated quantum code migration. However our work focuses on Qiskit which is a library for Python programming language. There is evidence that suggest LLMs may perform differently by language~\cite{twist2025llmslovepythonstudy}.

The work by Sahoo et. al \cite{sahoo_systematic_2024} provides a very complete review on prompt engineering and the most relevant associated works, considering query techniques, contextual refinement, and precision on the associated metrics, as well as prompt iteration and rephrasing.  Recent prompting innovations such as Rephrase-and-Respond (RaR) by Deng et al. \cite{deng_rephrase_2024} and Chain-of-Thought prompting  \cite{Kojima-llms_zero-shot} have shown measurable benefits in task performance across ambiguous or complex queries. We test these prompting strategies under the added stressor of Qiskit version drift and demonstrate that their effectiveness degrades in the absence of domain adaptation—highlighting new boundary conditions for prompt engineering in quantum code tasks.

We also consider recent work related to Qiskit migration tasks \cite{ziftci_migrating_2025}. While it shares some overlap with our focus—particularly in emphasizing the manual difficulty, time cost, and error-prone nature of migrating quantum code—it takes a different approach by leveraging an indexing system (Kythe) to manage code references. Although it also classifies migration scenarios and uses the Gemini model, its evaluation does not take place within a complex quantum computing context. Instead, it relies heavily on human intervention for edge cases and final validations. Furthermore, the effectiveness of its technique is tightly coupled to the model's context window, whereas our approach is independent of such limitations, as the base taxonomy is precomputed and model-agnostic.

While prior work has established the potential of LLMs in code explanation, refactoring, and generation, there remains a substantial gap in their application to API migration in quantum software, especially within fast-evolving ecosystems like Qiskit. Our study addresses this need by benchmarking LLMs on their ability to explain and adapt code in the face of version drift. By combining model comparisons, explanation quality evaluation, and prompt variation experiments, we offer practical insights into LLM readiness for version-aware quantum software support.

\section{Taxonomy of Qiskit Migration Scenarios}
To effectively guide large language models (LLMs) in the task of Qiskit code migration, we rely on a structured taxonomy of migration scenarios. This taxonomy captures the changes introduced between Qiskit versions. These taxonomies need to be created per version release. I.e.: the taxonomy for version 0.46 is based on the change logs, documentation and release notes for that Qiskit version. The taxonomy was specifically designed to bridge the gap between unstructured release notes and actionable migration knowledge for automated refactoring systems.

\begin{figure}[h]
    \centering
    \includegraphics[width=.9\linewidth]{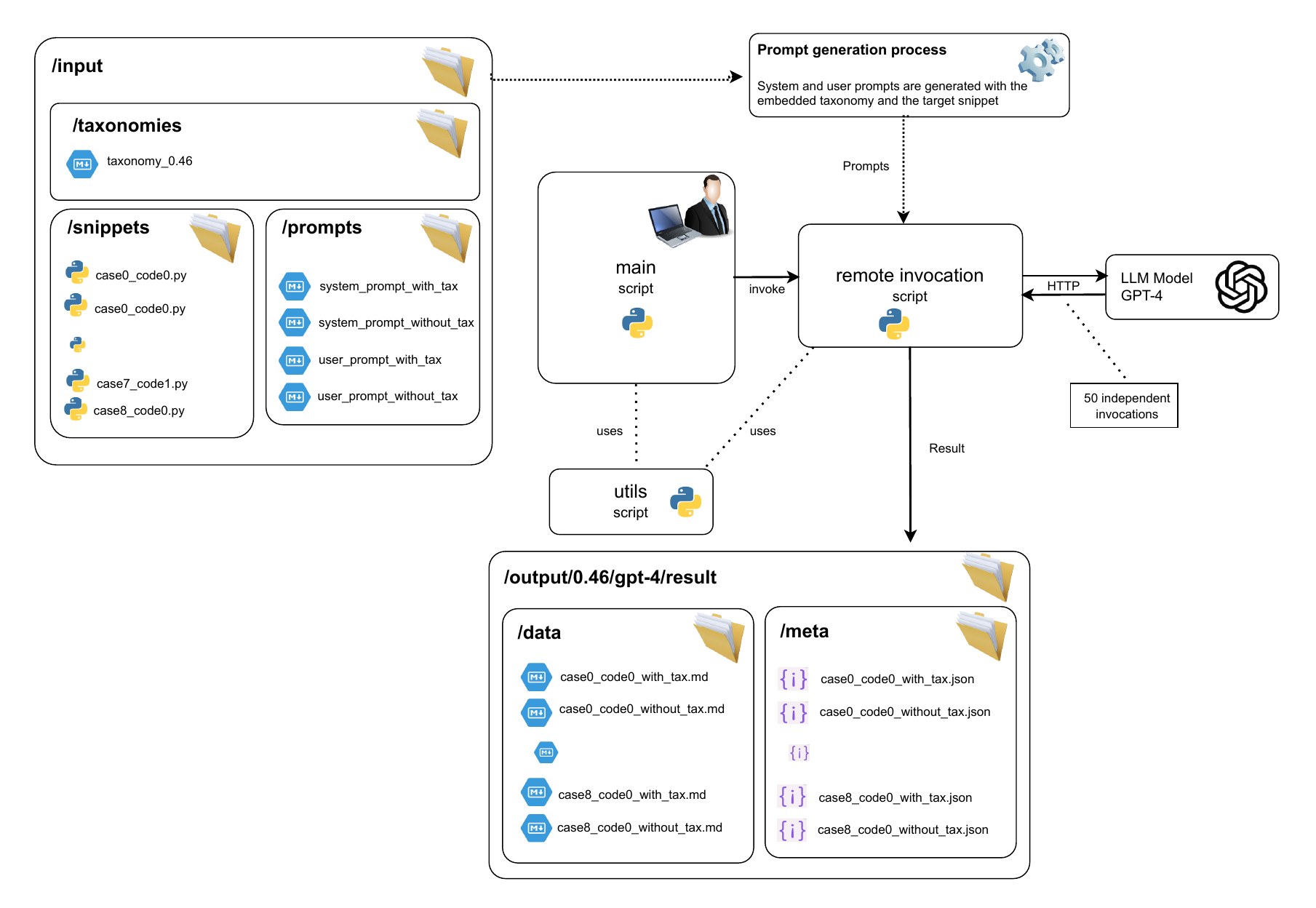}
    \caption{Architecture of the code repository used for the experiments. There are user and system prompts for both experiments, with and without taxonomy. The were a total of 50 invocations to the model, two per each of the 25 code snippets.}
    \label{fig:flow_diagram}
\end{figure}

The process involved in the creation of these taxonomies is described in detail in  Suárez et al.~\cite{previous_paper}. In a nutshell, the taxonomies can be created manually or using LLM assisted technologies. In both cases the origin of the migration scenarios is the official Qiskit documentation and release notes. The architecture of the code repository is described in Figure~\ref{fig:flow_diagram}.

While the work described above focuses on asses the ability of LLMs to summarize the changes between Qiskit versions in a structured manner and automatically generate the taxonomy, the focus of our study is to evaluate the impact of explicitly providing such structured information to LLMs in order to assist quantum software practitioners in identifying and performing code migrations.

To assess the influence of the taxonomy on the LLM-assisted migration process, we focused on synthetic yet representative migration cases targeting Qiskit version 0.46, constructed from known changes introduced in prior versions. Thus, we created a taxonomy targeting version 0.46 using ChatGPT 4.1. The results were refined and manually verified by experts in quantum computing programming. 

The resulting taxonomy is a \textit{markdown} file with columns described in Table~\ref{tab:taxonomy-dimensions}.

\begin{table}
\centering
\begin{tabular}{|l|p{3cm}|}
\hline
\textbf{Column} & \textbf{Description} \\
\hline
\textbf{Category} & Type of each migratory scenario. \\
\hline
\textbf{Migration Flow} & Source and Target of the version upgrade associated with the scenario. \\
\hline
\textbf{Summary} & Brief description of the scenario. \\
\hline
\textbf{Artifacts} & Summary of keywords of the artifacts involved. \\
\hline
\textbf{Example code in source version} & Python example code in the source version of the flow or previous versions. \\
\hline
\textbf{Example code in target version} & Python example code in the target version of the flow or later. \\
\hline
\textbf{Degree of Difficulty} & Complexity weighting involved in the migration scenario. \\
\hline
\textbf{Degree of impact in SE/QSE} & Relationship of the migrating scenario to aspects of classical software engineering and/or association with quantum software engineering. \\
\hline
\textbf{References} & Links to the authoritative sources of the scenario, either primary or secondary. \\
\hline
\end{tabular}
\vspace{1em}
\caption{Each scenario is framed by the LLM using these dimensions according to the provided prompt. The degree of difficulty and impact on SE/QSE, while returned by the tool were not used in the present work as require more discussion.}
\label{tab:taxonomy-dimensions}
\end{table}

The generated taxonomy consist of 43 scenarios, of which 29 are deprecations, 6 describe new features and 8 describe structural changes.

\section{Methodology}
To evaluate the impact of a structured taxonomy on the effectiveness of large language models (LLMs) in assisting Qiskit code migration, we designed a controlled experiment using synthetic yet representative code samples. A total of 25 Python code snippets were manually crafted to emulate realistic usage patterns while embedding specific migration scenarios relevant to Qiskit version 0.46. Each snippet consisted of between 9 to 30 lines  of code and was designed to reflect real-world practices such as varied import styles, inline comments, and multi-line constructs. All snippets targeted deprecated features or modules, or were constructed to expose the model to known patterns from the taxonomy. The number of cases per source file is between one and two but the associated lines to address a scenario may differ. For example, for the deprecation of the function \texttt{execute()} from \texttt{qiskit} module, the source code contains a line for the import sentence and the usage sentence in which the function is used. In this case the scenario is the same but multiple lines are affected.

For each code sample, we conducted two independent interactions with the same LLM: one where the full migration taxonomy was included as context, and one where it was omitted. In both cases, the prompts asked the model to identify migration issues and provide a structured response detailing the affected lines, a brief categorization description of the scenario, the artifact involved, and the suggested refactoring. When the taxonomy was included, the prompt additionally required the model to reference the corresponding scenario identifier or indicate with an "*" if no match was found, indicating the model found a migration scenario not present in the taxonomy. Line numbers were pre-inserted in the code through an automated preprocessing step to improve referential clarity in the output. To standardize the model outputs and facilitate comparison, we instructed the LLM to return its migration suggestions in a structured markdown table format. The expected output varied slightly depending on whether the taxonomy was provided. Notice the case in which no taxonomy is provided, in which is impossible to associate a scenario ID. 

In the case with taxonomy, the output was structured using six columns:
    \begin{itemize}
      \item \textbf{Line} — the code line number, added programmatically before the prompt.
      \item \textbf{Code} — the exact line of source code being analyzed.
      \item \textbf{Scenario ID} — the identifier of the migration scenario in the taxonomy, or an asterisk (\texttt{*}) if no match was found.
      \item \textbf{Scenario} — a synthesized description combining the taxonomy’s \emph{Type} and \emph{Summary} fields (e.g., \emph{Deprecation → function\_name() deprecated}). If the update is not mandatory for compatibility, the label \emph{(optional)} was added.
      \item \textbf{Artifact} — the affected component, drawn directly from the taxonomy’s \emph{Artifacts} field.
      \item \textbf{Refactoring} — the recommended code change for versions $\geq$ 0.46, left blank if the model was unsure or no clear fix was applicable.
    \end{itemize}

In the case without taxonomy, the output was structured using five columns:
    \begin{itemize}
      \item \textbf{Line} — the code line number.
      \item \textbf{Code} — the exact source code line.
      \item \textbf{Scenario} — a short description of the change and the affected artifact (e.g., \emph{Deprecation → function\_name() method deprecated}), including the \emph{(optional)} label if applicable.
      \item \textbf{Artifact} — the module, method, or parameter involved in the migration.
      \item \textbf{Refactoring} — the proposed code change for compatibility with version $\geq$ 0.46.
    \end{itemize}

All experiments were conducted using OpenAI's \texttt{gpt-4-0613} model via the \texttt{chat.completions} API endpoint, with a temperature setting of 0.1 and default values for all other parameters. No chain-of-thought prompting, function calling, or external tools were used. Furthermore, the model was not allowed internet access during the experiments. Although the prompts explicitly stated that the model could rely on its prior knowledge (the phrase "using your prior knowledge" was included into the prompt), we recognize that the specific versions of Qiskit included in its training corpus are not transparent. As such, we assume the model had access to documentation up to Qiskit versions 0.46, 1.0, and possibly 2.0 at the time of training, which introduces an uncontrollable variable discussed later in the discussion.

\begin{figure}[h]
    \centering
    \includegraphics[width=.9\linewidth]{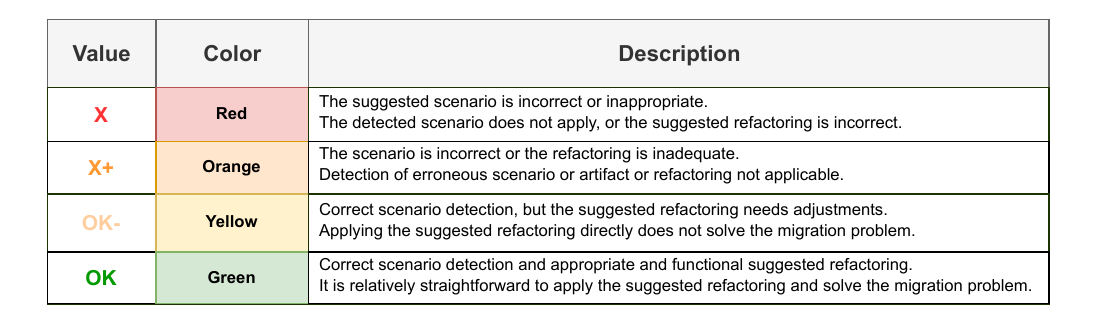}
    \caption{Evaluation criteria for the refactoring suggestions provided by the model. The same criteria was used for both, with taxonomy and without taxonomy.}
    \label{fig:metrics_figure}
\end{figure}

To assess the quality of the responses, two quantum computing experts manually evaluated each model suggestion. Each individual line-level recommendation was scored using a color-coded rubric: green for correct suggestions (OK); yellow for minor issues easily fixed on inspection (OK-); orange for valid Python but not aligned with version 0.46 or not adapted to the provided code (X+); and red for incorrect or misleading suggestions or when the scenario is misidentified (X). The review process was performed independently without formal inter-rater reliability measures, though all scores were cross-checked for consistency. Also, in order to account for the amount of false negatives, the number of not reported migration changes was verified. We provide these numbers in the result section as a \textit{ratio of missed changes} over the total expected changes. See Figure~\ref{fig:metrics_figure}.

This evaluation framework allowed us to measure whether prompting with a structured taxonomy leads to better migration outcomes than relying solely on the LLM’s pre-trained knowledge. Ultimately, our goal is to determine whether the overhead of building a structured taxonomy is justified by tangible improvements in LLM-assisted migration support.

\section{Results}
We evaluated the impact of providing a structured taxonomy on the performance of a large language model (LLM) when identifying and assisting with migration scenarios in Qiskit code. The evaluation focused on two key dimensions: scenario identification and the quality of refactoring suggestions.

\subsection*{Scenario Identification}

A total of 25 synthetic code samples were used in the experiments. Of these, 21 contained scenarios that required code refactoring due to incompatibility with Qiskit version~0.46. The remaining 4 were already compatible and served as negative test cases.

Across the 21 samples that required refactoring, a total of 81 lines of code were expected to be modified.

\begin{itemize}
  \item With the taxonomy, the model correctly identified 12 of the 21 refactoring scenarios.
  \item Without the taxonomy, 10 of the 21 scenarios were correctly identified.
  \item In both experiments, only 1 out of the 4 non-refactoring scenarios was correctly recognized as such (true negative).
  \item In the remaining 3 non-refactoring cases, the model produced incorrect migration suggestions (false positives).
\end{itemize}

These results indicate a moderate improvement in scenario identification when the taxonomy is included in the prompt.

\subsection*{Refactoring Suggestions}

We also assessed the quality of the refactoring suggestions provided by the model for the 81 lines of code that required changes.

\begin{itemize}
  \item When using the taxonomy, the model correctly refactored 50 out of 81 lines (true positives).
  \item Without the taxonomy, the model produced correct suggestions for only 29 out of 81 lines.
  \item The taxonomy-assisted model produced 40 incorrect or misleading changes (false positives), compared to 61 incorrect suggestions without the taxonomy.
  \item The model failed to detect or provide incorrect suggestions on 31 lines that required refactoring when guided by the taxonomy, while it missed 52 of such lines in the absence of the taxonomy (false negatives).
\end{itemize}

These results suggest that access to a structured taxonomy significantly improves both the precision and recall of the LLM in migration tasks. The taxonomy helps reduce the number of incorrect suggestions while also enabling the model to identify a higher number of correct refactorings.

All scripts, associated data and results are contained in our GitHub repository and are fully accessible without restriction~\cite{sofware-experiments}.

The results achieved, described above, are summarized in Table~\ref{tab:results_summary}.


\begin{table}[ht]
\centering
\renewcommand{\arraystretch}{1.2} 
\setlength{\tabcolsep}{2pt}       
\begin{tabular}{|p{4cm}|c|c|c|}
\hline
\textbf{Metric} & \textbf{w/Tax} & \textbf{wo/Tax} \\
\hline
\multicolumn{3}{|l|}{\textbf{Scenario Identification}} \\
\hline
Correctly identified scenarios that need refactoring (TP) & 12  & 10  \\
Correctly identified scenarios that does not need refactoring (TN) & 1   & 1   \\
FP & 9   & 11   \\
FN & 3   & 3   \\
\hline
Precision & 0.57 & 0.47\\
Recall & 0.85 &  0.76\\
\hline
\multicolumn{3}{|l|}{\textbf{Refactoring Suggestions}} \\
\hline
Correct suggestions (TP)  & 50  & 29  \\
Incorrect suggestions (FP) & 40  & 61  \\
Missed suggestions (FN)  & 31  & 52  \\
\hline
Precision & 0.55 & 0.32\\
Recall & 0.62 & 0.35\\
\hline
\end{tabular}
\vspace{0.8em}
\caption{Summary of scenario identification and refactoring results with and without taxonomy support.}
\label{tab:results_summary}
\end{table}

\section{Discussion}
Our work shows the role of the taxonomy as a condensed representation of migration scenarios. This structured format can support developers in adapting their code by enabling large language models (LLMs) to more effectively identify migration instances. Because the taxonomy is significantly more compact than the raw documentation from which it is derived, it allows for a greater portion of the model’s context window to be allocated to actual code, without the need for techniques such as Retrieval-Augmented Generation (RAG). While comparing our approach to alternatives—such as including unprocessed documentation in the prompt or employing retrieval-based augmentation—is beyond the scope of this work, it is worth noting that the documentation associated with a single version can easily exceed the context length of contemporary LLMs.

Our work only focuses on the migration to a single version of Qiskit from code compatible with the same minor releases (i.e: from 0.41-0.45 to 0.46).  This approach aligns with the recommendation in the Qiskit documentation, which advocates for incremental version migration to ensure greater reliability and maintainability. Migration across non-consecutive versions, particularly in the case of major version changes, introduces additional complexity and requires a more rigorous and individualized analysis.

Another possible avenue of investigation involves prompting the model to directly produce a migrated version of the input code. Although our primary goal was to evaluate the effectiveness of the taxonomy in helping the model to identify migration scenarios, applying the same evaluation framework to the refactored output code proved more complex and was therefore not adopted in this study. Nevertheless, our methodology could be adapted for such an evaluation in future work, and exploring the impact of the taxonomy on the quality of migrated code remains a compelling direction.

Given the observed benefits and the growing prevalence of LLM-assisted coding tools, it may be advisable for maintainers of the Qiskit library to consider appending structured taxonomies to their release notes. This official source of taxonomies may prevent the blooming of different not official taxonomies. In our observations, improvements in the quality and completeness of the taxonomy led to better performance by the model in specific cases—particularly when documentation was ambiguous or incomplete. However, the impact of taxonomy quality on model performance was not quantitatively assessed in this study and warrants further investigation.

Finally, prompt engineering also emerges as a relevant factor. In our case, the prompts did not explicitly instruct the model to adapt the refactoring to the specific context or variable names of the input code. As a result, we observed occasional inconsistencies, such as the reuse of placeholder variables copied verbatim from the taxonomy examples. Refining the prompt design to guide the model more precisely in contextual adaptation is an open opportunity for optimization. For example, relax or tight the prompt to allow or disallow flexibility in the responses.

Regarding the refactoring dimension of the experiment, the results reveal a nuanced balance in the quality of the suggestions provided by the model. In some instances, the model produced overly simplistic recommendations that failed to address the underlying migration issue, often requiring additional searches and manual corrections. For example, in cases involving deprecated module imports, the model occasionally suggested their direct removal without considering their subsequent usage within the code, when a proper replacement would have been more appropriate. Conversely, in other cases, the proposed refactorings proved problematic due to poor adaptation to the specific context of the code, either by being too generic or misaligned with the intended functionality. Moreover, some suggestions exhibited an excessive level of specificity or were poorly suited to the actual migration need, ultimately failing to solve the compatibility issue.

These findings suggest that achieving an effective balance between precision and flexibility in code migration tasks may depend heavily on the clarity and specificity of the prompt provided to the model—commonly referred to as prompt engineering~\cite{sahoo_systematic_2024}. Furthermore, improvements in the comprehensiveness and granularity of the taxonomy used to guide the model are also likely to enhance the quality and applicability of the refactoring recommendations.

It is also worth mentioning that this study focused solely on the GPT-4 model (specifically `gpt-4-0613`). Evaluating other models—especially those designed with a stronger orientation toward code understanding and generation, such as Codex, CodeLlama, or Claude for code—could provide valuable insights into the generalization of our findings. Exploring how different architectures handle taxonomy-based guidance remains an interesting and promising line of future work.

All in all, our results suggest that the use of a taxonomy is useful to guide the process of identifying migration scenarios as well as to inform potential refactoring suggestions. At the same time, we observed that the model is able to effectively complement the structured information provided by the taxonomy with its own prior knowledge, as inferred from the prompt instructions. The refactoring suggestions were, in many cases, well adapted to the specific context of the input code. 

The experiments confirm the model’s accuracy in detecting the lines of code that require refactoring and in referencing the corresponding scenario from the taxonomy (column \textit{Scenario Id}). While the model typically analyzes code on a line-by-line basis, it also shows the ability to identify and coherently group logically related segments. On the other hand, we observed that the model does not attempt to identify refactoring scenarios that lie outside the target version explicitly specified in the prompt. In our study, this target was version 0.46.0, and the model consistently scoped its responses to that version.

We regard this experimental study as part of a broader line of inquiry, building upon our previous work (Suárez et. al~\cite{previous_paper}) aimed at understanding the key factors that influence the accuracy of large language models (LLMs) in the context of quantum software refactoring. This includes identifying the strengths of current approaches as well as areas requiring improvement, with the goal of outlining the scope and practical applicability of LLMs for quantum software engineering (QSE) tasks.

In particular, we plan to enhance the structure, categorization, and supporting metadata used in the automatic generation of migration taxonomies. We hypothesize that the quality of this process will have a direct impact on both the precision of scenario identification and the relevance of the refactoring recommendations generated by LLMs. In addition, we envision the development of complementary tools focused on explainability and actionable insights, tailored to the specific improvements introduced in each new Qiskit release. These tools would support software teams in maintaining compatibility with cutting-edge versions while mitigating technical debt.

Furthermore, we aim to advance the development of impact metrics that quantify the effect of refactoring in quantum software, in alignment with the metadata previously integrated into our taxonomy. Such metrics will allow for a more systematic assessment of the efficiency gains derived from updated algorithms, the adoption of new framework features, and the transparency of code modifications. As part of future work, we intend to define new metrics and refine existing ones to better capture the trade-offs involved in the refactoring process.

Finally, we plan to extend the current evaluation pipeline to include recently released versions of Qiskit, particularly version 2.0, for which it is certain that the model lacks training data. This extension will allow us to assess model performance in scenarios beyond its training cutoff, offering valuable insights into its generalization capacity and adaptability to unseen migration requirements.

\bibliographystyle{plain}
\bibliography{biblografia}

\end{document}